\documentclass[]{aa}

\usepackage{savesym}
\usepackage{amsmath}
\savesymbol{iint}
\usepackage{txfonts}
\restoresymbol{TXF}{iint}

\usepackage{graphicx}
\usepackage{natbib}
\usepackage{epsfig}

\begin{document}

\title{A new equilibrium torus solution and GRMHD initial conditions} \titlerunning{Equilibrium
  torus solution} 
  \author{Robert F. Penna \inst{1}, 
  Akshay Kulkarni, \and Ramesh Narayan \inst{2}}
  \institute{
  $^{1}$Department of Physics, and Kavli Institute for Astrophysics and Space Research,
Massachusetts Institute of Technology, Cambridge, MA 02139, USA\\
  $^{2}$Harvard-Smithsonian Center for Astrophysics, 60 Garden
  Street, Cambridge, MA 02138, USA\\ \email{rpenna@mit.edu}
  (RFP)
%\\ \email{rnayaran@cfa.harvard.edu} (RN)
}
\date{\today}

\abstract
%Context
{General relativistic magnetohydrodynamic (GRMHD) simulations are providing influential models for black hole spin measurements, gamma ray bursts, and supermassive black hole feedback.  Many of these simulations use the same initial condition: a rotating torus of fluid in hydrostatic equilibrium.   A persistent concern is that simulation results sometimes depend on arbitrary features of the initial torus.  For example, the Bernoulli parameter (which is related to outflows), appears to be controlled by the Bernoulli parameter of the initial torus.}
%Aims
{In this paper, we give a new equilibrium torus solution and describe two applications for the future.  First, it can be used as a more physical initial condition for GRMHD simulations than earlier torus solutions.  Second, it can be used in conjunction with earlier torus solutions to isolate the simulation results that depend on initial conditions.}
%Methods
{We assume axisymmetry, an ideal gas
  equation of state, constant entropy, and ignore self-gravity. We fix an angular momentum
  distribution and solve the relativistic Euler equations in the
  Kerr metric.}
%Results
{The Bernoulli parameter, rotation rate, and geometrical thickness of
  the torus can be adjusted independently.  Our torus tends to
  be more bound and have a larger radial extent
  than earlier torus solutions.}
%Conclusions
{While this paper was in preparation, several GRMHD simulations appeared based on our equilibrium torus.  We believe it will continue to provide a more realistic starting point for future simulations.}


\maketitle

\section{Introduction}

Accretion flows onto black holes are typically magnetized and turbulent, so general relativistic magnetohydrodynamic (GRMHD) simulations have played an influential role in model building.  The initial conditions for many of these simulations are the same: a rotating torus of fluid held together by gravity, pressure gradients, and centrifugal forces \citep{fishbone76,fishbone77,kozlowski78,
abramowicz78,chakra85}.  The entropy and angular
momentum distribution of the torus are chosen arbitrarily and then hydrostatic equilibrium and an equation of state 
fix the fluid's density, velocity, and pressure. At the start of a simulation, the
magnetorotational instability (MRI) 
\citep{balbus91,balbus98} develops and the torus becomes turbulent.  Turbulence transports angular momentum outward, allowing the fluid to accrete inwards, and the inner edge of the torus becomes an accretion flow.  The torus typically persists throughout the simulation and provides a reservoir of fluid feeding the outer edge of the accretion flow.

Simulations based on the equilibrium torus solutions of \citet{fishbone76} and \citet{chakra85} have found many applications: thin disk models \citep{2008ApJ...687L..25S,2009ApJ...692..411N,penna2010} black hole spin evolution \citep{2004ApJ...602..312G}, radio emission from Sgr A* \citep{2007CQGra..24S.259N,2012MNRAS.426.1928D,2012ApJ...755..133S}, black hole jets \citep{2006MNRAS.368.1561M,2005astro.ph..6369M,2009ApJ...704..937N,sasha2011,2012MNRAS.423L..55T}, computations of spectra  \citep{2010MNRAS.401.1620H}, magnetized accretion with neutrino losses \citep{2011NewA...16...46B,2007PThPh.118..257S,2008AIPC.1054...79B}, pair production in low luminosity galactic nuclei \citep{2011ApJ...735....9M},  numerical convergence studies \citep{2012ApJ...744..187S}, tilted disk evolution \citep{2007ApJ...668..417F,2008ApJ...687..757F,2012ApJ...761...18H}, binary black hole mergers  \citep{2011PhRvD..84b4024F,2012PhRvL.109v1102F}, and magnetically arrested disks \citep{devilliers2003,mckinney2012}.

\begin{figure}
\begin{center}
\includegraphics[width=0.99\columnwidth]{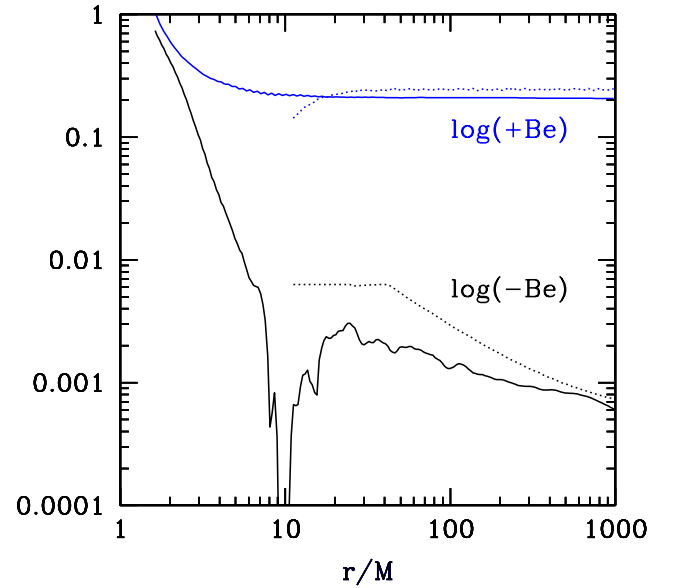}
\end{center}
\caption{The initial (dotted) and final (solid) midplane Bernoulli
  parameters, $Be$, of two GRMHD accretion simulations.  The Bernoulli parameter is measured in units of rest mass energy.   The
  energetically unbound flow (blue) is the A0.0BtN10 model of
  \citet{mckinney2012}.  The energetically bound flow (black) is the
  ADAF/SANE model of \citet{narayan12}.  The unbound simulation reached $t=96,796M$ and the bound simulation reached $t=200,000M$. In both
  cases, the final $Be$ appears to depend on the initial
  $Be$. }
\label{fig:bes}
\end{figure}

The equilibrium tori of \citet{fishbone76} and \citet{chakra85} are designed to be simple.  They assume unphysical, power law angular momentum distributions in order to keep the solutions analytical.  When they are used as the initial condition for GRMHD simulations, one hopes the turbulent accretion flow ``forgets'' unrealistic features of the initial torus.  However, this does not always seem to be the case.  For example, the Bernoulli parameter of the initial torus appears to persist through to the final accretion flow (Figure \ref{fig:bes}).

The Bernoulli parameter is the sum of the kinetic energy, potential energy, and enthalpy of the gas (at least in Newtonian dynamics, where this splitting can be made precise.  See \S\ref{sec:BeMHD} for a discussion of the GRMHD Bernoulli parameter.)  At large distances from the black hole, the potential energy vanishes.  Since the other two terms are positive, gas at infinity must have $Be \geq 0$.  Furthermore, in steady state and in the absence of viscosity, $Be$ is conserved along streamlines.  Hence any parcel of gas that flows out with a positive value of $Be$ can potentially reach infinity.  A flow with positive $Be$ is called unbound and a flow with negative $Be$ is called bound.  Unbound flows are more likely to generate outflows.

As shown in Figure \ref{fig:bes}, the Bernoulli parameter of the accretion flows in some GRMHD simulations appears to be set by the initial torus. This is a concern, as it suggests the strength of simulated outflows, such as winds, might be sensitive to arbitrary choices in the initial conditions.   GRMHD simulators should choose the initial $Be$ with some care.  For example, the true Bernoulli parameter of an accretion flow onto a supermassive black hole is probably related to its value at the outer edge of the accretion flow, where $Be\sim 0$.  So it would be reasonable to simulate this flow with an initial torus with $Be\sim 0$.  However, in the earlier solutions of \citet{fishbone76} and \citet{chakra85}, the Bernoulli parameter is tied to other parameters of the torus, such as its thickness, which one would like to vary independently.  The solution in this paper is more flexible and allows for varying the Bernoulli parameter and torus thickness independently.

%The initial torus of \citet{mckinney2012} is an energetically unbound cylinder
%extending radially outward to infinity.  Our equilibrium torus
%solution is energetically bound at scale heights for which the
%earlier solutions are unbound.  So as a GRMHD initial condition, it
%will likely produce weaker (or non-existent) winds.  Equilibrium tori
%with intermediate Bernoulli parameter are of course possible.  
%\citet{mckinney2012} studied accretion disk winds with GRMHD
%simulations.  In their simulations, the initial equilibrium torus is
%thick (scale height $|h/r|\sim 0.9$) and has a positive Bernoulli
%parameter.  In units of the gas rest mass energy, it has the dimensionless value $Be \sim 0.25$.  

The Bernoulli parameter is not the only arbitrary feature of GRMHD initial conditions that may affect the final results.  It is known that GRMHD simulation results depend on the initial magnetic field strength and topology \citep{beckwith2008,penna2010,sasha2011,mckinney2012}.    The rotation rate of the initial torus may also be important. For example, simulations which
start from slowly rotating tori will tend to be more convectively unstable than
simulations which start from rapidly rotating tori, as rotation
tends to stabilize accretion flows against convection. 

Our paper is organized as follows.  In \S\ref{sec:solution}, we construct our equilibrium torus
solution.  In \S\ref{sec:prop}, we obtain approximate analytical formulae for the radius of the outer edge, the
radius of the pressure maximum, the Bernoulli parameter, and the
geometrical thickness of the torus.    In \S\ref{sec:conc}, we summarize our
results.  In the Appendix, we describe a magnetic field
configuration for the torus and discuss the Bernoulli parameter of the
magnetized torus.  The magnetic field consists of multiple poloidal
loops and is constructed so that the magnetic flux and gas-to-magnetic
pressure ratio are the same in each loop.  This setup could be useful
for GRMHD simulations.

\section{New torus solution}
\label{sec:solution}

We consider a fluid torus in hydrostatic equilibrium
around a Kerr black hole.  We assume the flow is axisymmetric and stationary.  We use an ideal gas equation of state and assume the fluid has a constant entropy. Self-gravity is ignored.

We work in Boyer-Lindquist coordinates $\left(t,r,\theta,\phi\right)$
and units for which $G=c=1$.
The solution is a function of $r$ and $\theta$ only (i.e., it is stationary and axisymmetric).  There are six adjustable parameters.  They are defined below but may be summarized here: $r_{in}$ is the radius of the inner edge of the torus, $r_1$ and $r_2$ control the shape of the angular momentum distribution, $\xi$ controls the rotation rate, $\kappa$ sets the entropy, and $\Gamma$ is the adiabatic index.

%compute_gd
 The
components of the Kerr metric we need are
\begin{align}
g_{tt} &= -1 + 2Mr/\Sigma, \quad
g_{t\phi} = -2 M a r \sin^2\theta / \Sigma, \\
g_{\phi\phi} &= \left[r^2 + a^2 
           \left( 1 + 2Mr\sin^2\theta / \Sigma \right)\right] \sin^2 \theta,
\end{align}
where $M$ and $a$ are the mass and spin of the black hole and $\Sigma=r^2 + a^2 \cos^2\theta$.

%von Zeipel cylinders
Let $u^\mu$ be the fluid four-velocity. We assume the fluid's angular momentum density, $\ell\equiv u_\phi /|u_t|$, is constant on von Zeipel cylinders
\citep{abramowicz71}.  The radius, $\lambda$, of a von Zeipel
cylinder with angular momentum $\ell$ is \citep{chakra85}:
\begin{equation}\label{eq:lambda}
\lambda^2 = - \ell(\lambda) \frac{\ell(\lambda) g_{t\phi} + g_{\phi \phi}}
                                 {\ell(\lambda) g_{tt}+g_{t\phi}}.
\end{equation}
In Newtonian gravity, $\lambda = r\sin{\theta}$, the usual cylindrical
radius.  In the Schwarzschild metric, $\lambda =
\sqrt{-g_{\phi\phi}/g_{tt}}$.  In the Kerr metric, one must solve equation \eqref{eq:lambda} numerically.

%lK
The angular momentum density of circular, equatorial
orbits in the Kerr metric is \citep{NT73}
\begin{equation}\label{eq:ellK}
\ell_K(r) \equiv u_\phi/|u_t| = \sqrt{Mr} \thinspace \mathcal{F}/\mathcal{G},
\end{equation}
where,
\begin{align}
\mathcal{F} &= 1 - 2 a_* / r_*^{3/2} + a_*^2 / r_*^2, \quad
\mathcal{G} = 1- 2/r_* + a_*/r_*^{3/2},\notag\\
r_* &= r/M, \quad a_* = a/M.\notag
\end{align}
%angular momentum distribution
We choose the angular momentum distribution of the fluid torus to be:
\begin{equation}\label{eq:ell}
\ell(\lambda) = 
\begin{cases}
\xi \ell_K(\lambda_{1}) &\mbox{if } \lambda < \lambda_{1} \\
\xi \ell_K(\lambda) &\mbox{if } 
        \lambda_{1} < \lambda < \lambda_{2} \\
\xi \ell_K(\lambda_{2}) &\mbox{if } \lambda > \lambda_2.
\end{cases}
\end{equation}
There are three regions.  In the inner and outer regions of the torus, the angular momentum density is a constant independent of radius.  The size of these regions is set by $\lambda_1$ and $\lambda_2$.  At intermediate radii, the angular momentum density is a fraction $\xi$ of the Keplerian distribution \eqref{eq:ellK}.  This is chosen to be close to the expected, sub-Keplerian angular momentum distribution of a real accretion flow.  It is more physical than the power law angular momentum distributions of the \citet{fishbone76} and \citet{chakra85} solutions. 

The angular velocity of the
torus is
\begin{equation}\label{eq:Omega}
\Omega \equiv \frac{u^\phi}{u^t}
       = -\frac{g_{t\phi}+\ell g_{tt}}
           {g_{\phi\phi}+\ell g_{t\phi}}.
\end{equation}
We have assumed $u^r=u^\theta=0$, so this fully determines the velocity.

%Computations at torus inner edge
Given the velocity of the torus, we can determine its density and pressure from the Euler equation.  Let 
\begin{equation}
A \equiv u^t
= \left(-g_{tt}-2\Omega g_{t\phi}-\Omega^2 g_{\phi\phi}\right)^{-1/2},
\end{equation}
which sets the gravitational force felt by the fluid. The Euler equation  \citep{abramowicz78} is then
\begin{equation}\label{eq:euler}
\frac{\nabla p}{\rho_0+U+p}
 =\nabla \ln A 
 - \frac{\ell \nabla \Omega}{1-\Omega \ell}.
\end{equation}
  Our notation is standard: $\rho_0$, $U$, and $p$ are the mass density, internal energy, and gas pressure of the fluid in its rest frame.   Euler's equation describes the balance between pressure gradients (LHS) and gravitational and centrifugal forces (RHS) required for hydrostatic equilibrium.

%The redshift factor relates fluid frame
%proper time, $d\tau$, to Boyer-Lindquist time, $dt$, by $d\tau=dt/A$.
To solve the Euler equation, it is helpful to introduce the effective
potential \citep{kozlowski78}
\begin{equation}\label{eq:W}
W(r,\theta) \equiv - \ln(FA),
\end{equation}
where,
\begin{equation}\label{eq:F}
\ln F(r,\theta) \equiv- \int_{r_{in}}^{r_{\rm \lambda m}(r,\theta)} 
         {\frac{d\Omega}{dr}\frac{\ell dr}{1-\Omega\ell}}.
\end{equation} 
The lower limit of the integral, $r_{in}$, is the radius of the inner edge of the torus.  The upper limit, $r_{\lambda
m}(r,\theta)$, is the equatorial radius of the Von Zeipel cylinder
containing $(r,\theta)$.
%That is, $r_{\lambda m}$ is the Boyer-Lindquist radius that solves
%\eqref{eq:lambda} with $\theta=\pi/2$ for the current value of
%$\lambda$ and $\ell$.  
For example, $r_{\lambda m}(r,\pi/2)=r$.

In terms of the effective potential, the Euler equation is
\begin{equation}
\frac{\nabla p}{\rho_0+U+p} = -\nabla W.
\end{equation}
We can compute the effective potential because it depends only on $\ell$.  So the RHS is known. The boundary of the torus is the isopotential surface
$W(r,\theta)=W_{in}\equiv W(r_{in},\pi/2)$.  

The specific enthalpy is $w=1+\epsilon+p/\rho_0$, where $\epsilon=U/\rho_0$ is the specific internal energy.\footnote{We caution that $\epsilon$ is used for two different but related concepts in the literature.  In older papers, such as \cite{kozlowski78}, $\epsilon$ is the total energy density, $\rho_0+U$.  In more recent literature, such as \cite{devilliers2003}, $\epsilon$ is the specific internal energy, $U/\rho_0$.  We follow the latter convention.}   For an isentropic torus, the Euler equation can be integrated to obtain \citep{kozlowski78}
\begin{equation}
w(r,\theta)=e^{-\left(W(r,\theta)-W_{in}\right)}.\label{eq:w}
\end{equation}
We assume the equation of state $p=\rho_0 \epsilon(\Gamma-1)$, so the specific internal energy is
\begin{equation}
\epsilon = (w-1)/\Gamma.
\end{equation}
The rest mass density and pressure are
\begin{align}
\rho_0 &= \left[ (\Gamma - 1) \epsilon /\kappa
   \right]^{1/\left(\Gamma-1\right)}, \\
   p&=\kappa \rho_0^\Gamma \label{eq:U}.
\end{align}
The entropy, $\kappa$, is a free parameter.  The torus is now fully determined.

To summarize, we first choose the angular
momentum distribution \eqref{eq:ell}.   The angular momentum distribution determines the angular velocity and effective potential.  The effective potential determines the enthalpy of the torus through Euler's equation.  Fixing an ideal gas equation of state and assuming an isentropic torus, gives the density, pressure, and internal energy.  There are six free
parameters: the radius of the inner edge of the torus, $r_{in}$,
the break radii in the angular momentum distribution, $r_{1}$ and $r_{2}$, the normalization of the angular
momentum, $\xi$, the entropy, $\kappa$, and the adiabatic index, $\Gamma$.  The  entropy simply sets the density scale (equation \ref{eq:U}) and, as there is no self-gravity, it has no effect on the dynamics.

\section{Approximate analytical formulae}
\label{sec:prop}

The solution of \S\ref{sec:solution} can be implemented in GRMHD codes numerically.  However, for physical understanding, it is useful to have approximate formulae that describe the torus analytically.  In this section, we obtain approximate analytical formulae for the outer edge, 
pressure maximum, geometrical thickness, and Bernoulli
parameter of the torus.

The most complicated feature of the exact solution is the integral in equation \eqref{eq:F}.  To obtain approximate analytical formulae for the torus, we need to simplify this integral.  Let us first rewrite it as 
\begin{equation}\label{eq:F2}
F=\left(1-\Omega \ell \right)
 \exp \left( \int_{r_{in}}^{r_{\lambda m}}
 \frac{\Omega}{1-\Omega \ell}\frac{d\ell}{dr}dr \right),
\end{equation}
where we have used $\ell d\Omega = d(\Omega \ell)-\Omega
d\ell$. 

In the inner region of the torus ($\lambda < \lambda_1$),
the angular momentum is constant and the integrand in equation \eqref{eq:F2}
vanishes.  So
\begin{equation}
F(r,\theta)=1-\Omega(r,\theta) \ell(r,\theta), \quad (\lambda < \lambda_{1}).
\end{equation}

In the outer region of the torus ($\lambda>\lambda_2$), we may approximate the integral by plugging the Newtonian formula $\ell_K(\lambda)=\sqrt{M \lambda}$ into equation \eqref{eq:ell}, and using the Newtonian angular velocity $\Omega(\lambda)=\ell(\lambda)/\lambda^2$.  Now integrating from $\lambda_1$ to $\lambda_2$, we obtain

\begin{equation}\label{eq:Fout}
F \approx \left(1-\Omega \ell \right)I,
 \quad (\lambda > \lambda_{2}),
\end{equation}
where,
\begin{equation}
I \equiv
\left(\frac{1-\xi^2/\lambda_{2}}{1-\xi^2/\lambda_{1}}\right)^{1/2}.
\end{equation}
Equation \eqref{eq:Fout} is approximate because we ignored special relativistic
contributions to $\ell$ and $\Omega$.  But these are small in the outer  regions of the torus.  We can use our analytical approximation of $F$ to obtain simple formulae describing the torus.

\subsection{Radius of the outer edge}

The boundary of the torus is the isopotential surface
$W(r,\theta)=W_{in}=W(r_{in},\pi/2)$.  We can simplify $W$ at the outer edge by using
a Newtonian description there.  The equation for the outer radius,
$r_{out}$, becomes
\begin{equation}
W_{in} \approx -\frac{M}{r_{out}} 
    + \frac{\xi^2 M r_{2}}{2 r_{out}^2}-\ln I.
\end{equation}
The solution for the outer radius is
\begin{equation}
r_{out}/M \approx \frac{1+\sqrt{1- \xi^2\lambda_2\varpi}}{\varpi},
\end{equation}
where
\begin{equation}
\varpi \equiv 2 \ln \left[\frac{A}{I}
   \left(1-\frac{\ell^2}{\lambda^2}\right)\right]_{r=r_{in}}.
\end{equation}

\begin{figure*}
\begin{center}
\includegraphics[width=0.45\linewidth]{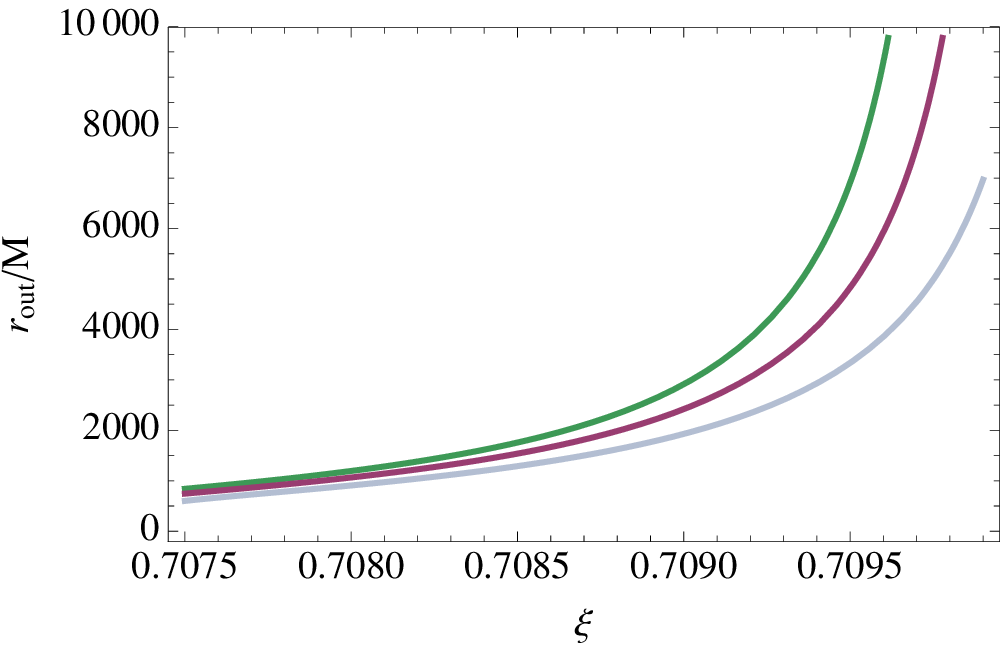}
\includegraphics[width=0.45\linewidth]{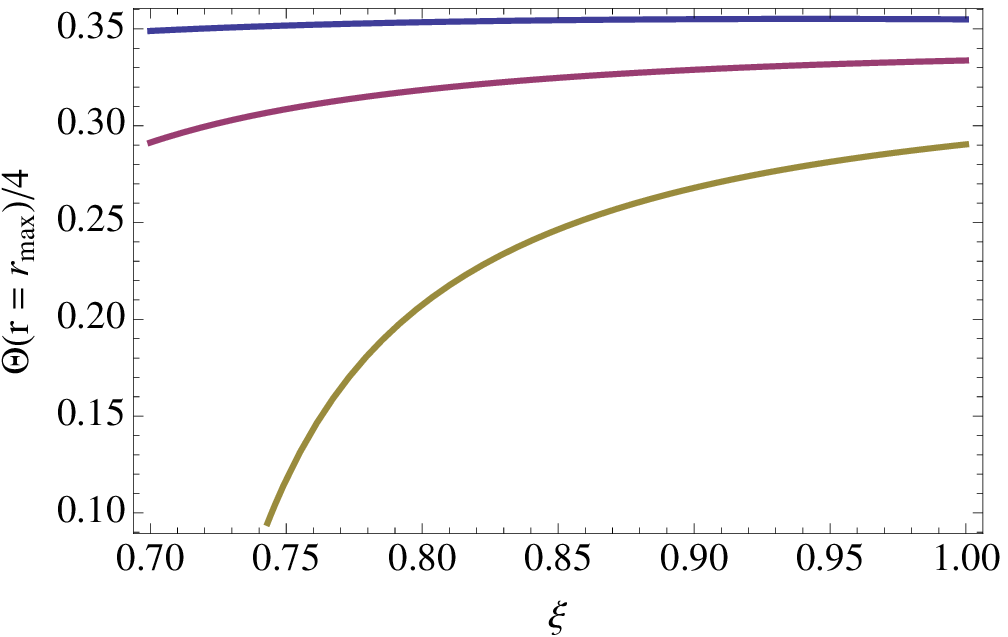}\\
\includegraphics[width=0.45\linewidth]{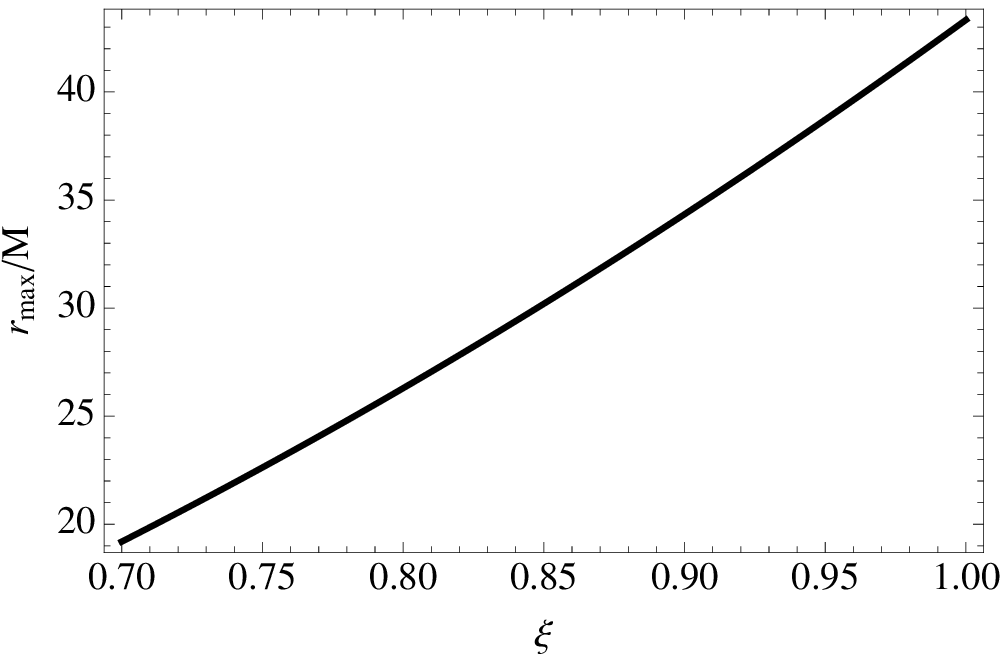}
\includegraphics[width=0.45\linewidth]{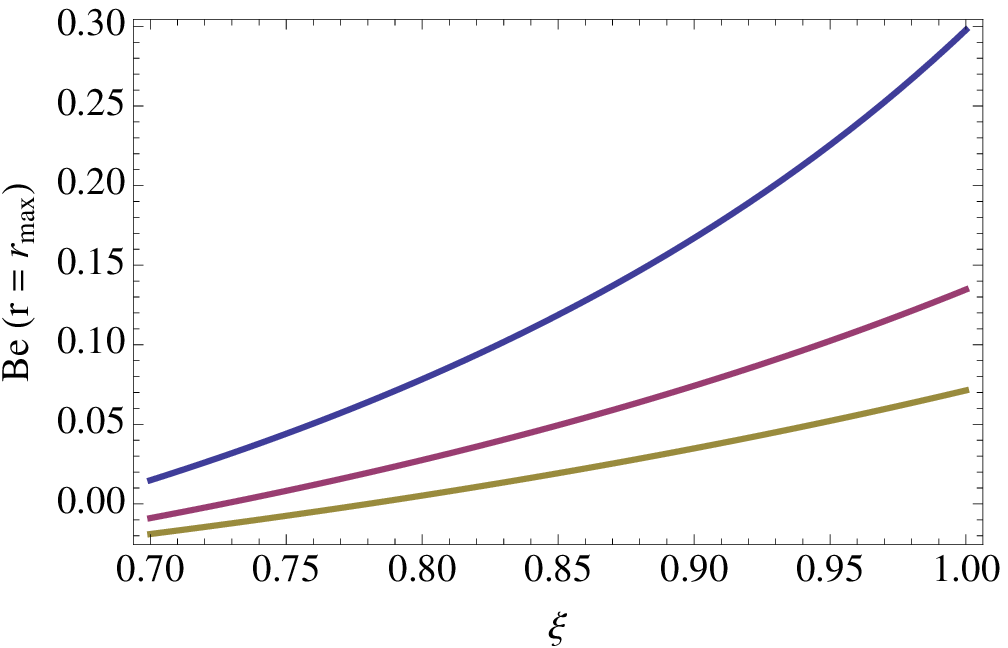}
\end{center}
\caption{Counterclockwise from the upper left panel, we show the
  radius of the outer edge of the torus, the radius of the pressure
  maximum, the Bernoulli parameter at the pressure maximum, and the
  geometrical thickness at the pressure maximum as a function of the
  rotation parameter, $\xi$, for several choices of $r_{in}/M$: 8
  (blue), 9.993 (green), 10 (red), 10.01 (gray), and 12 (yellow).  The
  pressure maximum position, $r_{max}/M$, is independent of $r_{in}$.
  The horizontal scale of the upper left panel is different from the
  other panels because $r_{out}$ is only finite when $Be<0$.}
\label{fig:props}
\end{figure*}

Figure \ref{fig:props} (top left panel) shows the variation of
$r_{out}$ with $\xi$ for several choices of $r_{in}$ and fixed
$r_1=42M$ and $r_2=1000M$.  The outer radius increases as the inner
radius decreases because the boundary of the torus is moving to larger
isopotential surfaces.  In the region of parameter space shown in
Figure \ref{fig:props}, the Bernoulli parameter is small and negative,
as we will see below.  When the Bernoulli parameter is small and
negative, the outer radius is very sensitive to $\xi$.  

Solutions with positive Bernoulli parameter have $r_{out}$ at infinity.  They are unbound and not very physical.  In fact, they are not even tori, they are infinite cylinders kept confined by a nonzero pressure
at infinity.  Unphysical, infinite cylinders based on the \citet{fishbone76} and \citet{chakra85} solutions have at times been used as the initial condition for GRMHD simulations.

\subsection{Pressure maximum}

The pressure gradient in the Euler equation \eqref{eq:euler}
is zero at the pressure maximum.  So the fluid must move along geodesics there.  In other words,
the pressure maximum is where
\begin{equation}\label{eq:rmaxcondition}
\ell(r_{max})=\ell_K(r_{max}),
\end{equation}
where $\ell_K$ is the angular momentum of circular, equatorial
geodesics (equation \ref{eq:ellK}) and $r_{max}$ is the radius of the pressure maximum.  For sub-Keplerian flow ($\xi<1$), the pressure
maximum must be in the inner region of the torus ($\lambda<\lambda_1$), because the angular momentum is strictly sub-Keplerian for $\lambda>\lambda_1$.  

Equation \eqref{eq:rmaxcondition} gives
\begin{equation}
r_{max}(\xi,\lambda_1) \approx \xi^2 \ell_K(\lambda_1)^2 - 4M,
\end{equation}
where $-4M$ is the leading order relativistic correction.
Figure \ref{fig:props} (lower left panel) shows the dependence
of $r_{max}$ on $\xi$ for $\lambda_1=42M$.  In the Keplerian limit $\xi\rightarrow 1$, the
pressure maximum approaches $\lambda_1$.  Lowering $\xi$ moves the pressure maximum toward the inner edge of the torus.

\subsection{Bernoulli parameter}
\label{sec:Be}

The relativistic Bernoulli parameter is \citep{NT73}
\begin{equation}\label{eq:Be}
%Be = -u_t \left(1+\frac{\Gamma U }{\rho_0}\right)-1.
Be = -u_t w - 1.
\end{equation}
Tori with $Be>0$ are energetically unbound and tori with $Be<0$ are
energetically bound.  In the inner region of the torus ($\lambda<\lambda_1$), the Bernoulli parameter is:
\begin{equation}
Be = -(g_{tt}+g_{t\phi}\Omega)
   \frac{A^2 (1-\Omega\ell)}
   {\left[A (1-\Omega\ell)\right]_{r = r_{in}}} -1.
\end{equation}
Figure \ref{fig:props} (lower right panel) shows how $Be(r_{max})$ depends on $\xi$ for several choices of
$r_{\rm in}$ and fixed break radii $r_1=42M$ and $r_2=1000M$.  The outer
edge of the torus goes to infinity as $Be\rightarrow 0$ (from below).
%  Note that by adjusting $r_{\rm in}$ and $\lambda_1$, any
%combination of $Be$ and $\xi$ is possible.

\subsection{Thickness}
\label{sec:thick}

The boundary of the
torus is the isopotential surface $W(r,\theta)=W_{in}$.  In the inner region of the torus ($\lambda<\lambda_1$), the
surface of the torus is
\begin{equation}
\sin\theta = -g_{tt} \frac{\ell^2}{r^2} 
  \frac{1}{1+g_{tt}\left[A\left(1-\Omega\ell\right)\right]_{r=r_{in}}}.
\end{equation}
$\theta$ is measured from the the polar axis, so the
 scale height of the torus is $\Theta =
\pi/2 - \theta$.  Figure \ref{fig:props} (upper right panel)
shows $\Theta(r_{max})$ as a function of $\xi$ for
several choices of $r_{in}$.

\subsection{Bernoulli parameter and thickness}
\label{sec:compare}

\begin{figure*}
\begin{center}
\includegraphics[width=0.45\linewidth]{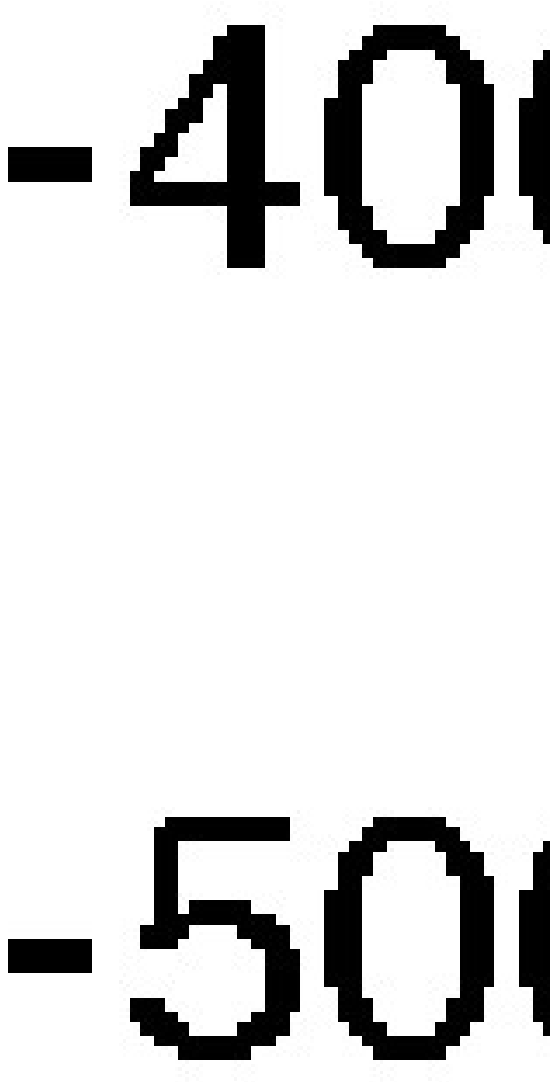}
\includegraphics[width=0.45\linewidth]{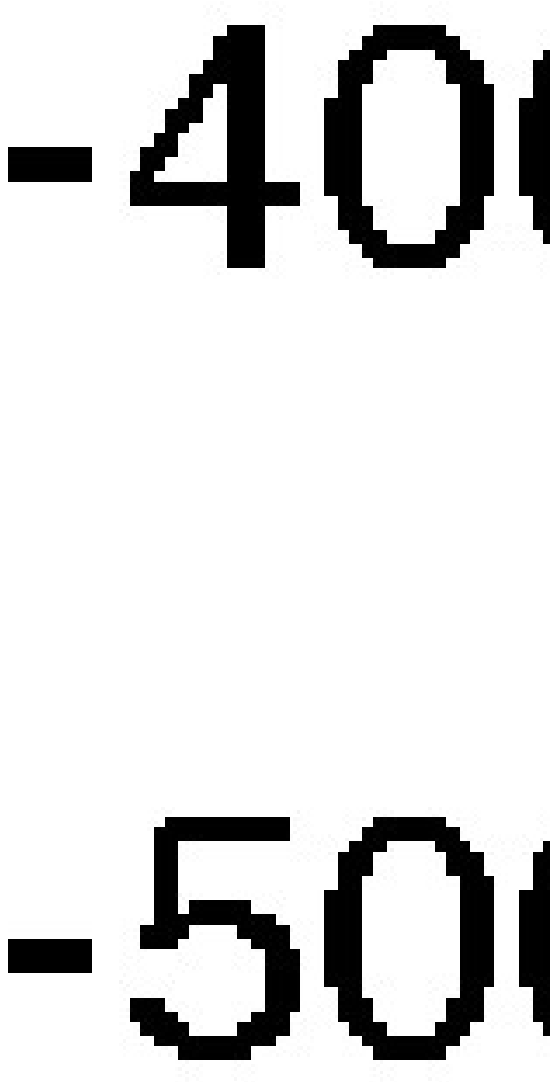}
\includegraphics[width=0.45\linewidth]{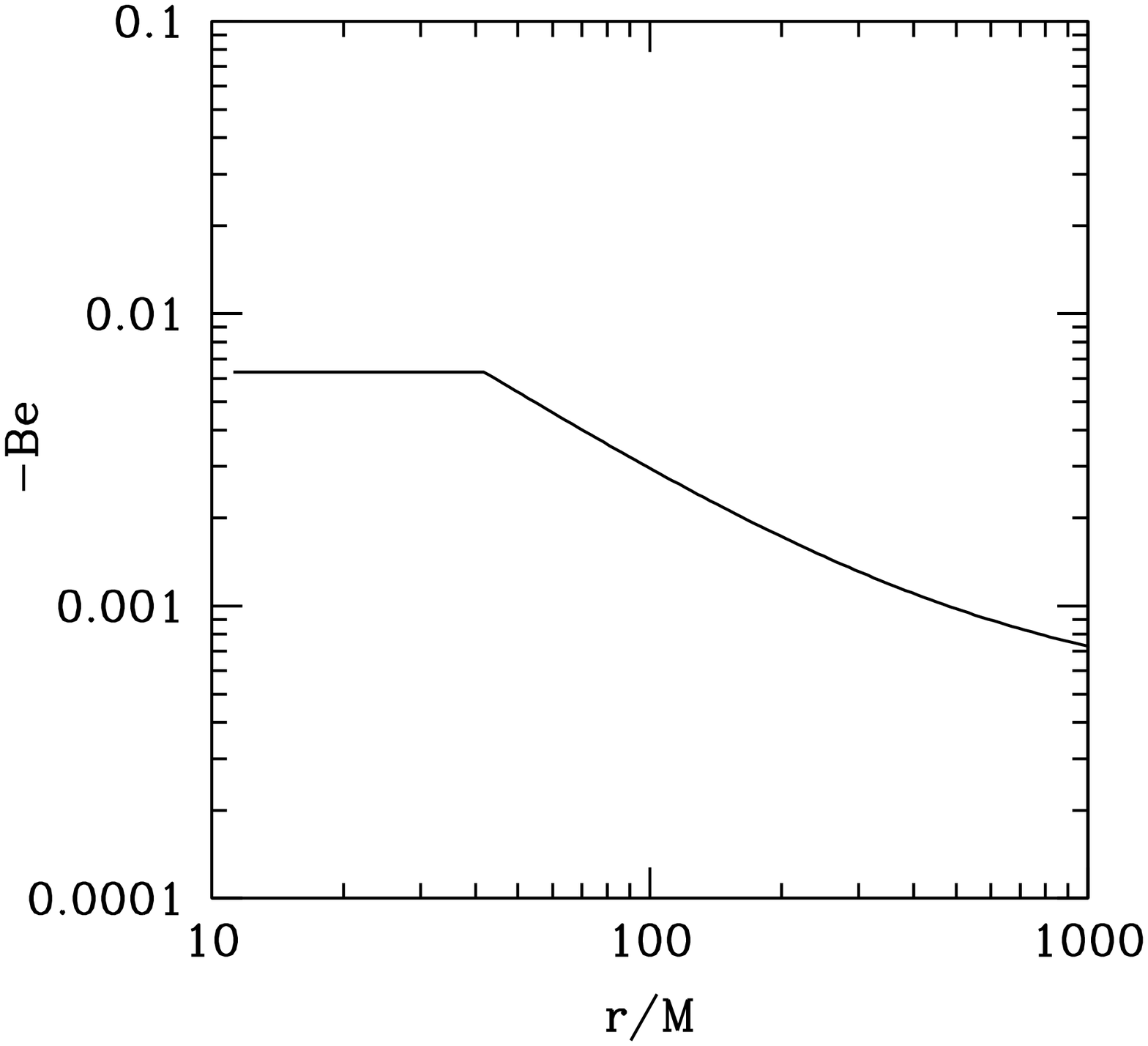}
\includegraphics[width=0.45\linewidth]{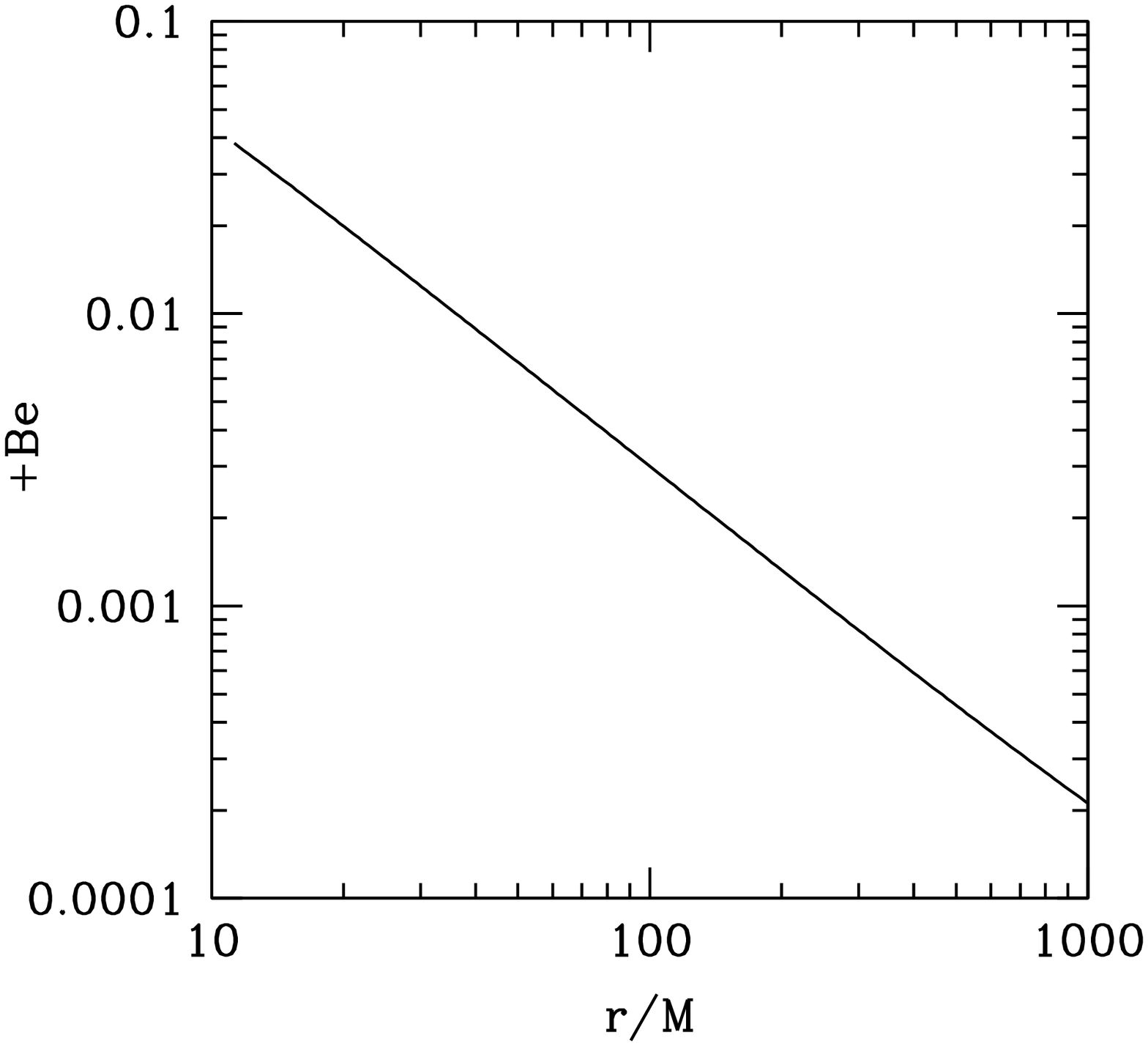}
\includegraphics[width=0.45\linewidth]{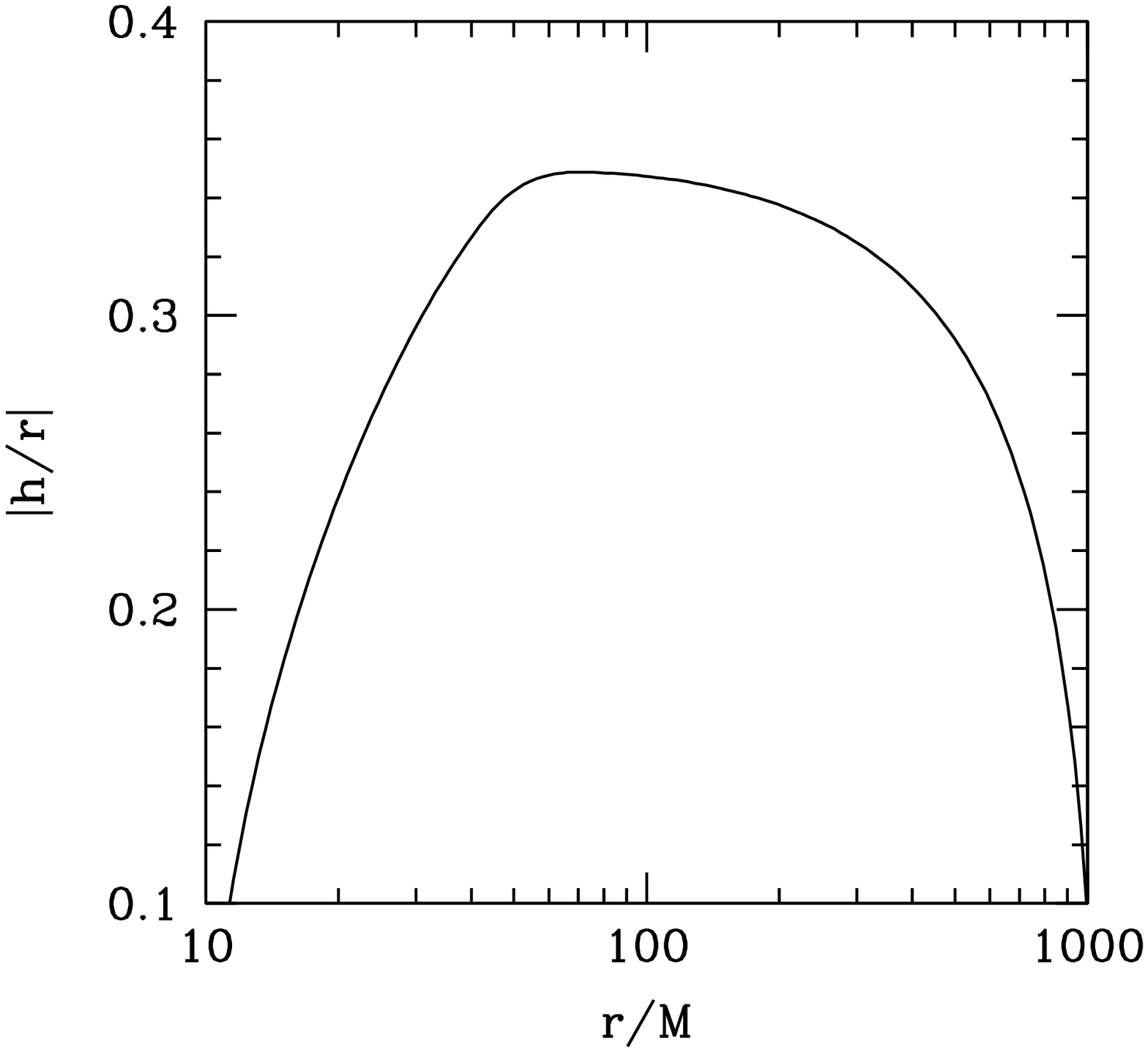}
\includegraphics[width=0.45\linewidth]{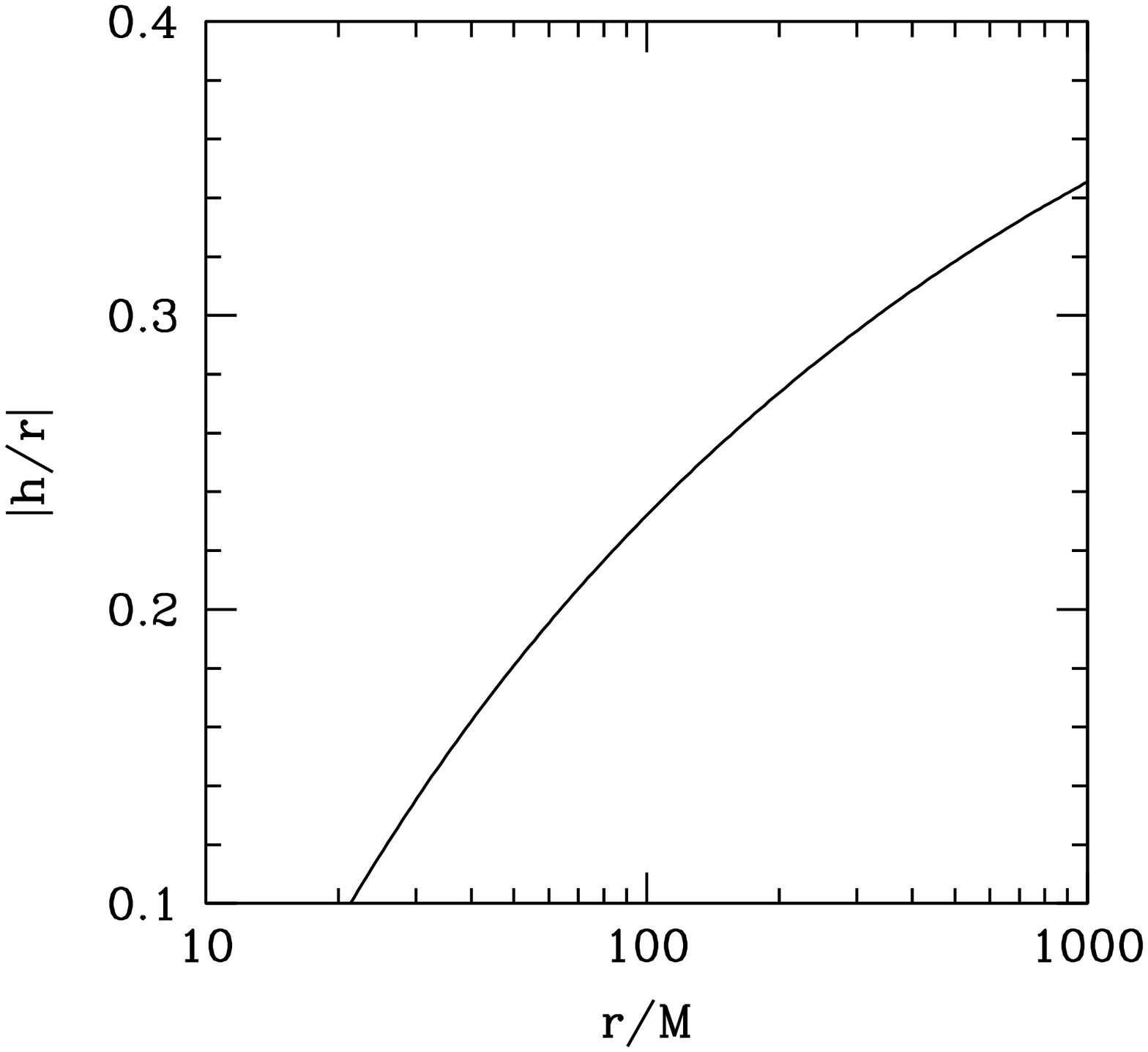}
\end{center}
\caption{The panels in the left column show $\log(-Be)$ and $|h/r|$
for the new equilibrium torus solution described in the text.  The
Bernoulli parameter is shown as a function of $(r,\theta)$ and as a
function of $r$ at the midplane.  The torus is energetically bound,
i.e., $Be < 0$.  The panels in the right column show $\log(Be)$ and
$|h/r|$ for a solution of \cite{chakra85} which is energetically
unbound, i.e., $Be > 0$.  Note that the unbound torus is actually thinner than the
bound torus in this example.}
\label{fig:Be}
\end{figure*}

Our solutions tend to be more bound than earlier solutions.  We show this with an example.  Figure \ref{fig:Be} (upper left panel) shows the Bernoulli parameter of one
of our solutions in the $(r,\theta)$ plane.\footnote{The free parameters are $r_{in}=10M$,
$r_{1}=42M$, $r_{2}=1000M$, $\kappa=0.00766$, and $\xi = 0.708$.}  The
torus is energetically bound: $Be<0$.  In the same figure (upper right panel), we show the
Bernoulli parameter of a solution of \citet{chakra85}.\footnote{The
free parameters are $r_{in}=10M$, $\kappa=0.00136M$, $d\log
\ell/d\log\lambda = 0.4$, and pressure maximum at $r_{\rm max}=40$.}
This torus is energetically unbound.

In the bottom panels of Figure \ref{fig:Be}, we show the density scale
heights,
\begin{equation}
|h/r| = \frac{\int \left| \theta-\pi/2 \right|\rho_0 \sqrt{-g}d\theta d\phi }
 {\int \rho_0 \sqrt{-g}d\theta d\phi },
\end{equation}
of both solutions.  Thinner tori tend to be more bound than thicker tori, but in this example the unbound
torus is actually thinner than the bound torus.  In other words, our solutions tend to be more bound than earlier solutions.  This is an advantage, because unbound solutions are unphysical as initial conditions for GRMHD simulations.  They are infinitely extended cylinders supported by pressure at infinity.

A second lesson to draw from this example is that the torus thickness, $|h/r|$, does not fix even the sign of the
Bernoulli parameter (much less its value).  GRMHD simulators typically focus on the thickness of the initial torus but this may not be enough if, for example, the Bernoulli parameter influences the strength of accretion disk winds.  In other words, it may be important to know the thickness and the Bernoulli parameter of the initial torus independently.

\section{Conclusions}
\label{sec:conc}

We have constructed a new equilibrium torus solution.  It provides an
alternative initial condition for GRMHD disk simulations 
to the earlier solutions of
\citet{fishbone76} and \citet{chakra85}.  The
angular momentum density is constant in the inner and outer regions of the
torus and 
follows a sub-Keplerian distribution in between.  The entropy is constant everywhere.

The Bernoulli parameter, rotation rate, and geometrical thickness of
the solution can be varied independently.  The torus tends to be more
energetically bound than earlier solutions, for
the same thickness $|h/r|$.  In fact, we have shown that it is possible to
generate equilibrium tori with the same $|h/r|$, but for which one is
bound and the other unbound.  So as GRMHD initial conditions, the new
solutions might lead to weaker (or non-existent) disk winds.  Future
GRMHD simulations will need to explore how the simulation results depend on the
Bernoulli parameter, rotation rate, and geometrical thickness of the
initial equilibrium torus.

In real accretion flows, the Bernoulli parameter is probably set at the outer edge of the flow.  If this is far from the black hole (for example, at the Bondi radius), then one expects $Be\sim 0$. It is not possible to obtain converged GRMHD simulations out to the Bondi radius.  One would need to run the simulations for a timescale of order the Bondi radius viscous time, which is impractical.    Computational resources limit the duration of even the longest run simulations to $t\sim 200,000 M$, which corresponds to the viscous time at $r\sim 200M$.  Probably the best one can hope to do is choose an initial condition with a realistic Bernoulli parameter ($Be\sim 0$) at the outset.  The solution in this paper is flexible enough to allow independent control over the thickness and Bernoulli parameter of the torus, so it is an ideal initial condition for GRMHD simulations.  In fact, while this paper was in preparation, there appeared several simulations based on our solution \citep{narayan12,2013arXiv1307.1143S,2013arXiv1307.4752P,2013MNRAS.428.2255P}.  We expect more soon.

%Radiatively inefficient accretion flows are expected to converge to a self-similar solution, at least away from the edges of the accretion flow.  The self-similar ADAF and CDAF solutions (refs) have Bernoulli parameters set by the outer edge of the flow.  One expects $Be \sim 0$ (ref).  The simulations, on the other hand, appear to maintain the Bernoulli parameters of the initial torus.  To see the $Be$ expected of the self-similar solutions, one probably has to resolve several decades in radius so that over a large range of radii the flow is unaffected by edge effects at the inner and outer boundaries of the flow.  In other words, one would like the simulations to converge out to $r=10^3-10^4$.  This will happen on a timescale of order the viscous timescale at these radii.  This is out of reach with the current generation of computing resources.  Someday it should become feasible.  In the meantime, one can use the torus solutions provided in this paper to generate initial tori which are tuned to have the expected $Be$.

\begin{acknowledgements}

R.F.P was supported in part by a Pappalardo Fellowship in Physics at MIT. 

\end{acknowledgements}

\begin{appendix}

\section{Adding a magnetic field}

Implementing the new equilibrium torus as a GRMHD initial condition
requires adding a magnetic field to the torus.  Here we record one
possible magnetic field, a series of poloidal magnetic loops.  We
discuss the Bernoulli parameter of the magnetized torus in
\S\ref{sec:BeMHD}.

\subsection{Magnetic field solution}
\label{sec:field}

We construct the magnetic field so that each loop carries the
same magnetic flux and $\beta = p_{\rm gas}/p_{\rm mag}$ is roughly
constant.  Simulations often require initial conditions that minimize
secular variability during the run, so these features can be useful.

Three free parameters appear in the solution: $r_{\rm start}$, $r_{\rm
end}$, and $\lambda_B$.  The first two set the inner and outer
boundaries of the magnetized region and the third controls the size of
the poloidal loops.

We define the field through the vector potential, $A_\mu$.  The
magnetic loops are purely poloidal, so only $A_\phi$ is nonzero.  To
keep $\beta$ close to a constant, the magnetic field strength
tracks the fluid's internal energy density.  Define
\begin{equation}
q=
\begin{cases}
\sin^3\theta\left(U_{c}/U_{cm}-0.2\right)/0.8 
   &\mbox{if } r_{\rm start}<r<r_{\rm end}\\
0  &\text{otherwise, }
\end{cases}
\end{equation}
where,
\begin{align}
U_{c}(r,\theta) &= U(r,\theta) - U(r_{\rm end},\pi/2), \\
U_{cm}(r) &= U(r,\pi/2) - U(r_{\rm end},\pi/2).
\end{align}
The function $q$ is defined to give $q=1$ at the midplane and
$q\rightarrow 0$ away from the midplane.  The factor $\sin^3\theta$
smooths the vector potential as it approaches the edges of the torus.
Dropping this factor leads to a torus with highly magnetized edges.

Further define
\begin{equation}
f(r) = \lambda_B^{-1} \left(r^{2/3} + 15 r^{-2/5}/8 \right).
\end{equation}
The vector potential is then
\begin{align}
A_\phi &= 
\begin{cases}
q \sin\left(f(r) - f(r_{\rm start})\right) &\mbox{if } q > 0\\
0  &\mbox{otherwise},
\end{cases}\\
 &\text{all other $A_\mu$=0.}
\end{align}
The sinusoidal factor in $A_\phi$ breaks the poloidal field into a
series of loops.  The function $f(r)$ gives each loop the same
magnetic flux.  The number of loops is controlled by $\lambda_B$.  The
overall normalization of $A_\phi$ has not been specified so it can be
tuned to give any field strength.

\begin{figure*}
\begin{center}
\includegraphics[width=0.45\linewidth]{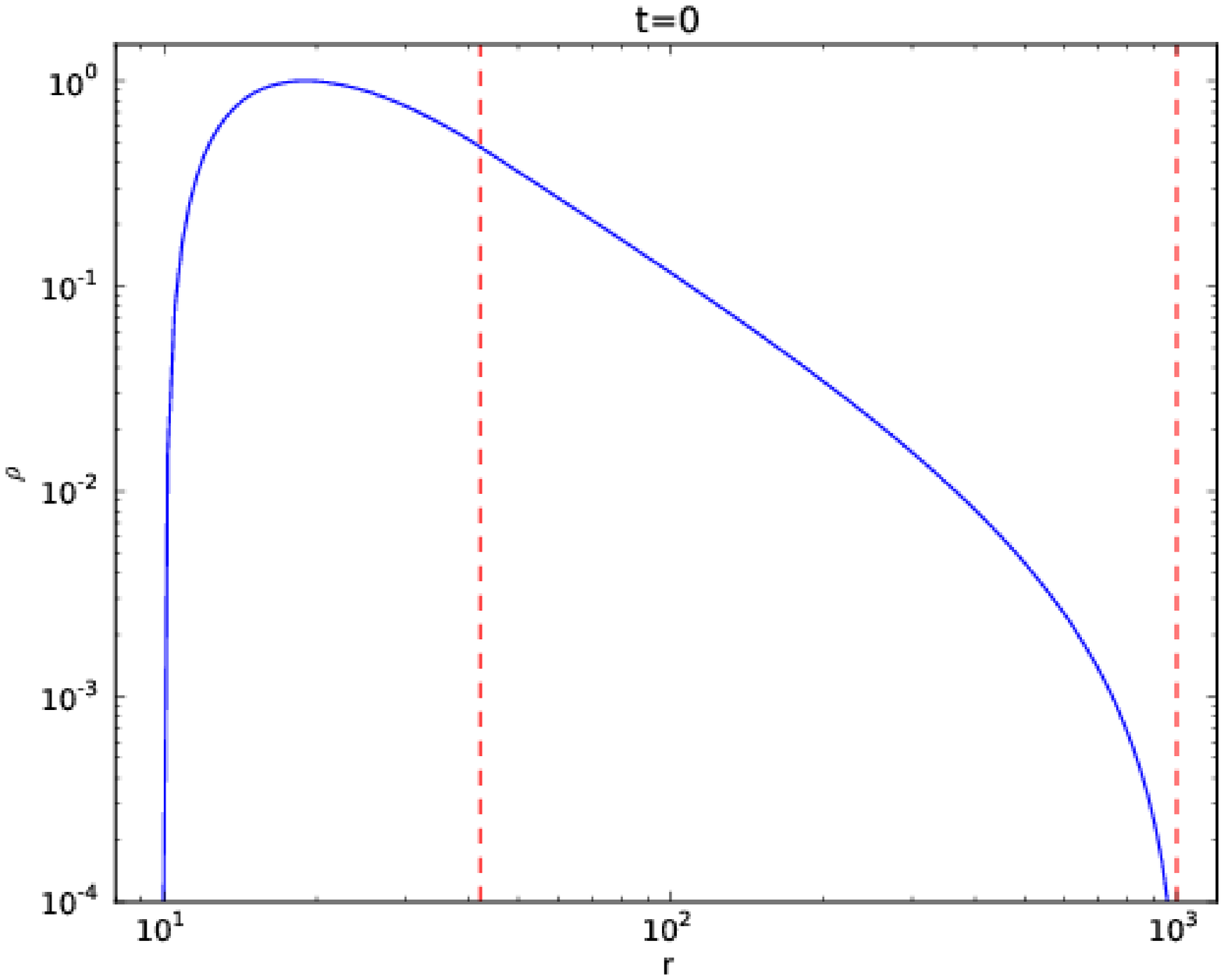}
\includegraphics[width=0.45\linewidth]{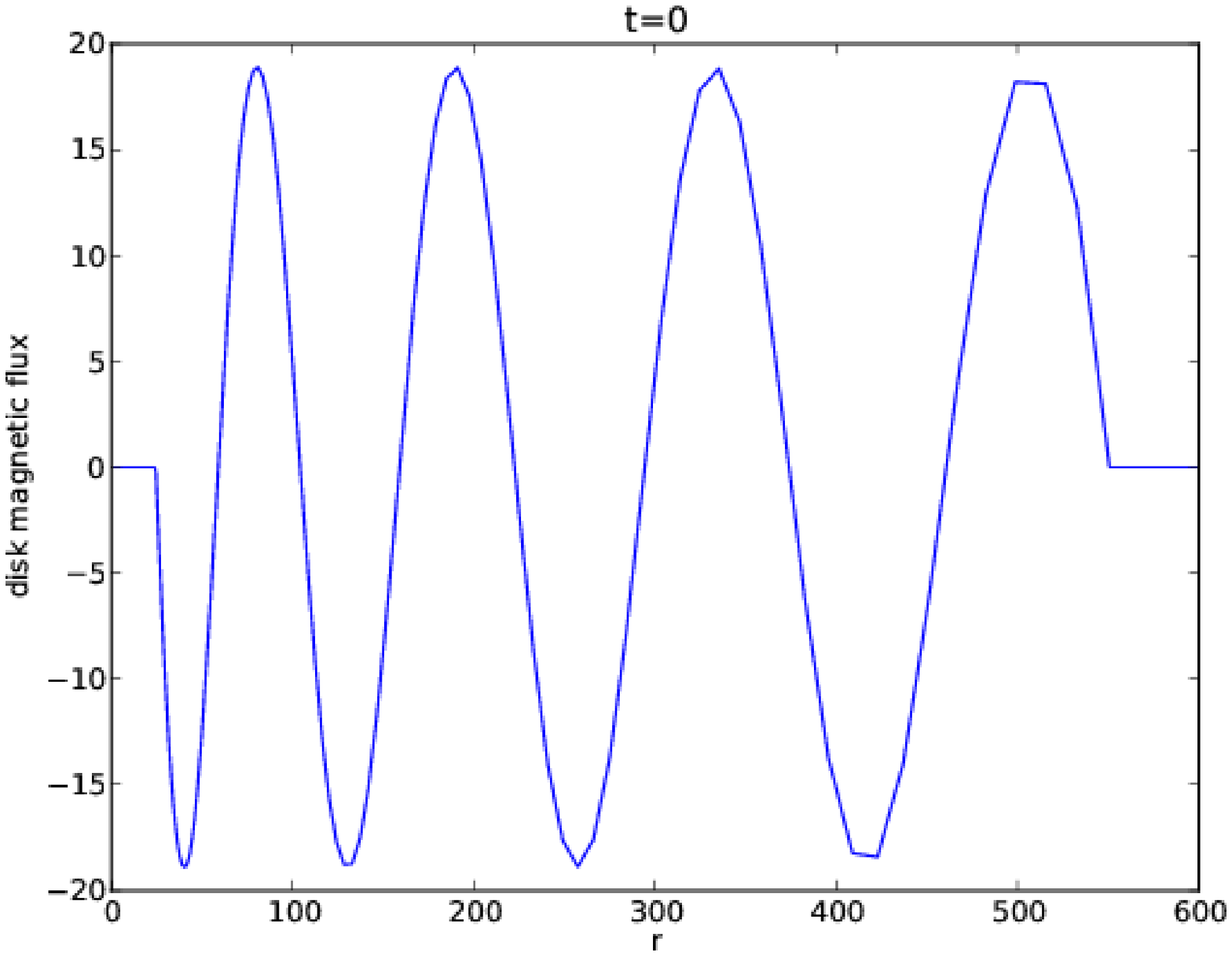}
\includegraphics[width=0.45\linewidth]{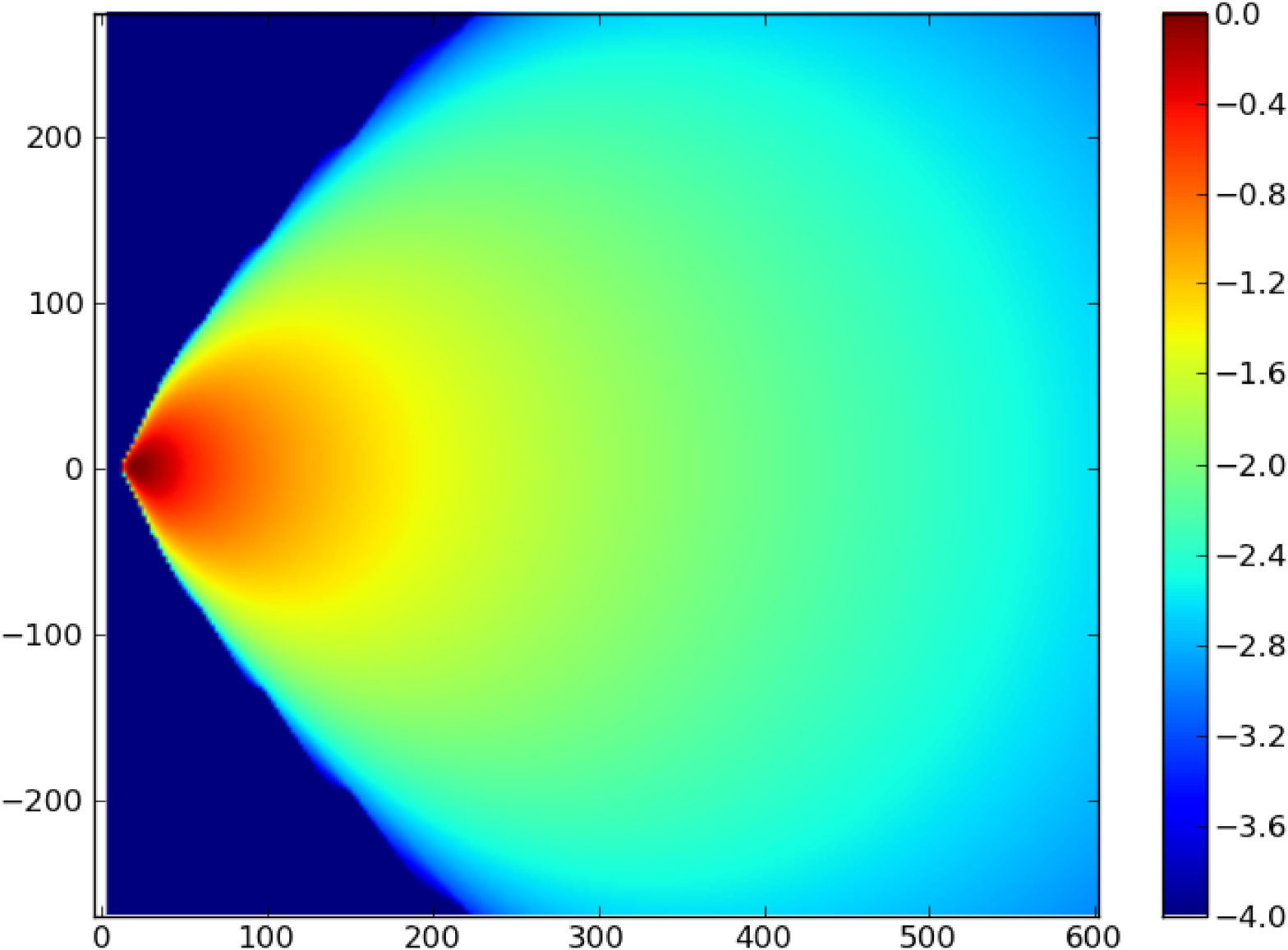}
\includegraphics[width=0.45\linewidth]{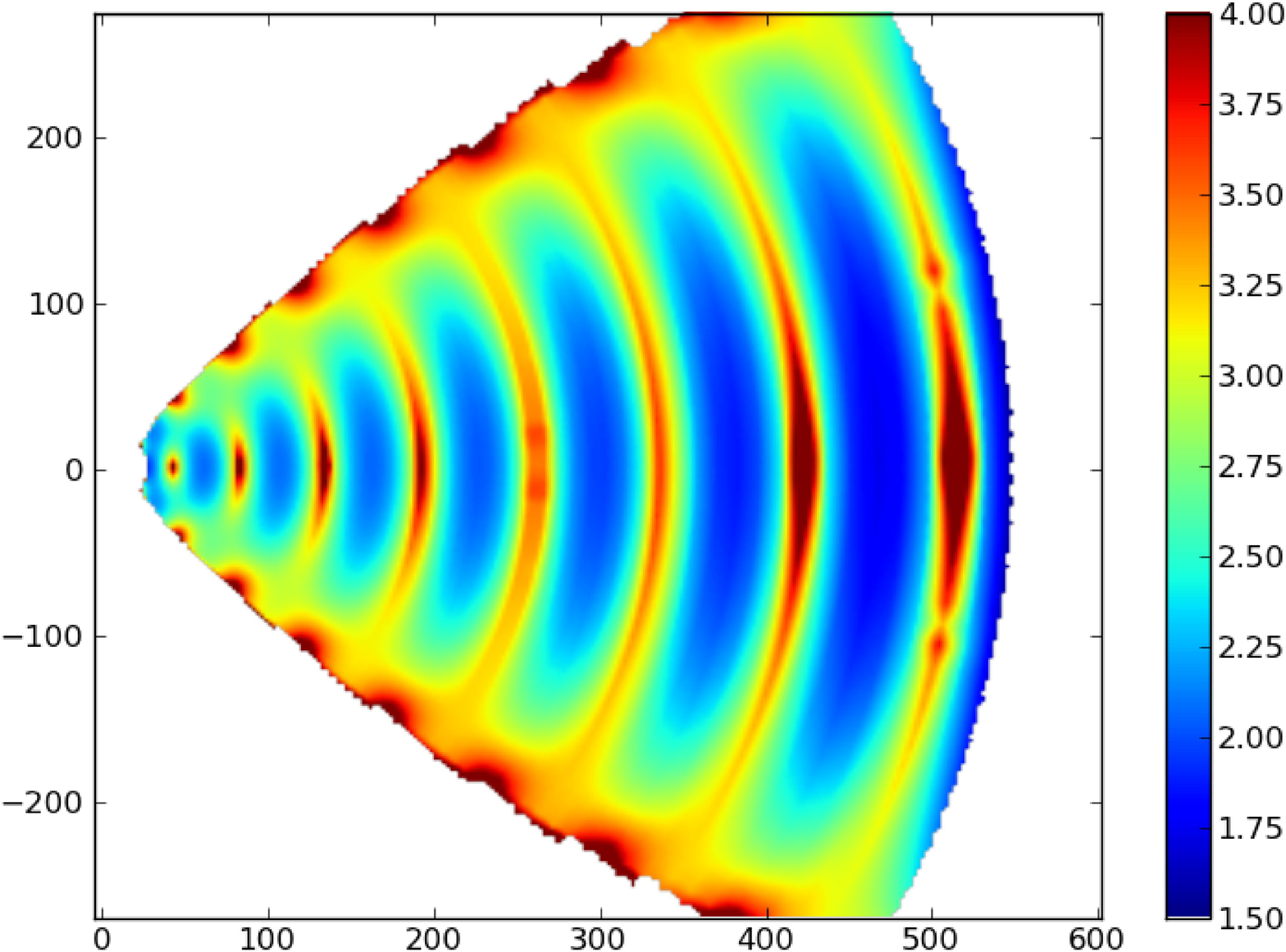}
\end{center}
\caption{The top two panels show the mid-plane density and the
  enclosed magnetic flux as a function of radius.  The lower two
  panels show the density and the magnetization parameter $\beta$ of
  the torus in the poloidal plane.  Each loop carries the same
  magnetic flux.  The magnetization is roughly constant throughout the
  torus.}
\label{fig:field}
\end{figure*}

We give an example in Figure \ref{fig:field}.  The equilibrium torus is
as in Figure \ref{fig:Be} and the field parameters are $r_{\rm
start}=25$, $r_{\rm end}=550$, and $\lambda_B=15/4$.  In this example
there are eight magnetic loops.  The magnetic flux, $A_\phi$, peaks at
the center of each loop, measures the flux carried by the loop, and is
the same across the torus.  The magnetization $\beta=p_{\rm gas}/p_{\rm mag}$
peaks at loop edges and drops at loop centers, but is roughly constant
across the torus.  In this example $\beta \sim 100$.

\subsection{GRMHD Bernoulli parameter}
\label{sec:BeMHD}

The stress energy tensor of a magnetized fluid is
\begin{equation*}
T_{\mu\nu}=\left(\rho_0+U+\frac{b^2}{2}\right)u_\mu u_\nu 
+\left(p_{\rm gas}+\frac{b^2}{2}\right)h_{\mu\nu}-b_\mu  b_\nu,
\end{equation*}
where $b_\mu$ is the magnetic field in the fluid's rest frame and
$h_{\mu\nu}=g_{\mu\nu}+u_\mu u_\nu$ is the projection tensor.  The
magnetic field contributes $b^2/2$ to the total internal energy and
$b^2/2$ to the total pressure, and introduces a stress term,
$-b_\mu b_\nu$.

The Euler equations, $h\cdot \left(\nabla\cdot T\right)=0$, become:
\begin{equation*}
\left(\rho_0+U+p_{\rm gas}+b^2\right)a=
-h\cdot\nabla \left(p_{\rm gas}+\frac{b^2}{2}\right)h
-h\cdot \left(b\cdot \nabla \right)b,
\end{equation*}
where $a=\nabla_u u$ is the fluid's acceleration.  We have used
$\nabla\cdot b = 0$ to simplify the last term on the RHS.

Assume the flow is stationary and adiabatic, project the Euler
equations along $\xi=\partial_t$, and combine terms using the first
law of thermodynamics.  This leads to:
\begin{equation*}
\frac{d}{d\tau}\left(\frac{\rho_0+U+p_{\rm gas}+b^2}{\rho_0}u_t\right)
=-\frac{1}{\rho_0}\xi\cdot h\cdot \left(b\cdot \nabla \right)b.
\end{equation*}
For the field configuration of \S\ref{sec:field}, $b$ is purely
poloidal and  $\xi\cdot h$ is purely toroidal.  So the RHS is zero.

We thus obtain a straightforward generalization of equation
\eqref{eq:Be}:
\begin{equation}\label{eq:bound}
Be=-\left(1+\frac{U+p_{\rm gas}+b^2}{\rho_0}\right)u_t -1.
\end{equation}
%Let us write $Be=\Phi+1/2 v^2 + w$, where $\Phi$ is the gravitational
%potential and $w$ is the total enthalpy (gas plus magnetic), and
%consider how magnetic fields change the Bernoulli parameter of the
%bound torus in Figure \ref{fig:Be}.  The unmagnetized torus has $\Phi
%+ 1/2 v^2 \sim - 2w_{\rm gas}$, so the Bernoulli parameter is $Be \sim
%-w_{\rm gas}$.  Adding magnetic fields to the torus, with
%gas-to-magnetic pressure ratio $\beta=p_{\rm gas}/p_{\rm mag}$, thus
%changes the Bernoulli parameter by $1/\beta$.  Simulations typically
%have initial $\beta\sim 100$, in which case the magnetic contribution
%to the Bernoulli parameter is of order $1\%$.
The unmagnetized torus has $Be \sim -w_{\rm gas}$.  Adding magnetic
fields to the torus, with gas-to-magnetic pressure ratio $\beta=p_{\rm
gas}/p_{\rm mag}$, changes the Bernoulli parameter by terms of order
$1/\beta$.  Simulations typically have initial $\beta\sim 100$, in
which case the magnetic contribution to the Bernoulli parameter is of
order $1\%$.

\end{appendix}

\bibliographystyle{mnras}
\bibliography{ms}

\end{document}